\documentclass[runningheads]{llncs}
\usepackage{multirow}
\usepackage{graphicx}
\usepackage{marvosym}
\usepackage{hyperref}
\usepackage{algorithm}
\usepackage{amsmath,amsfonts}
\usepackage{algpseudocode}
\usepackage{float}
\usepackage{diagbox}
\renewcommand{\arraystretch}{1.4}

\begin{document}
\title{GSMorph: Gradient Surgery for cine-MRI Cardiac Deformable Registration}
%
\titlerunning{Gradient Surgery for Deformable Registration}
%
%
\author{Haoran Dou\inst{1}\thanks{Haoran Dou and Ning Bi contributed equally to this work.} \and 
Ning Bi\inst{1\star} \and 
Luyi Han\inst{2,3} \and
Yuhao Huang\inst{4,5,6} \and
Ritse Mann\inst{2,3} \and
Xin Yang\inst{4,5,6,7} \and
Dong Ni\inst{4,5,6} \and
Nishant Ravikumar\inst{1,8} \and
Alejandro F. Frangi\inst{1,8,9,10} \and
Yunzhi Huang\inst{11}\textsuperscript{(\Letter)}
}
\authorrunning{Haoran Dou et al.}
%
%
%
\institute{
Centre for Computational Imaging and Simulation Technologies in Biomedicine (CISTIB), University of Leeds, Leeds, UK \and
Department of Radiology and Nuclear Medicine, Radboud University Medical Centre, Nijmegen, The Netherlands \and 
Department of Radiology, Netherlands Cancer Institute, Amsterdam, The Netherlands \and
National-Regional Key Technology Engineering Laboratory for Medical Ultrasound, School of Biomedical Engineering, Health Science Center, Shenzhen University, China \and 
Medical Ultrasound Image Computing (MUSIC) Lab, Shenzhen University, China \and
Marshall Laboratory of Biomedical Engineering, Shenzhen University, China \and
Shenzhen RayShape Medical Technology Co., Ltd, China \and
Division of Informatics, Imaging and Data Science, Schools of Computer Science and Health Sciences, University of Manchester, Manchester, UK \and
Medical Imaging Research Center (MIRC), Electrical Engineering and Cardiovascular Sciences Departments, KU Leuven, Leuven, Belgium \and
Alan Turing Institute, London, UK \and
Institute for AI in Medicine, School of Artificial Intelligence, Nanjing University of Information Science and Technology, Nanjing, China
\\
\email{yunzhi.huang.scu@gmail.com}}
\maketitle              
\begin{abstract}
Deep learning-based deformable registration methods have been widely investigated in diverse medical applications.
Learning-based deformable registration relies on weighted objective functions trading off registration accuracy and smoothness of the deformation field. Therefore, they inevitably require tuning the hyperparameter for optimal registration performance. Tuning the hyperparameters is highly computationally expensive and introduces undesired dependencies on domain knowledge. In this study, we construct a registration model based on the gradient surgery mechanism, named GSMorph, to achieve a hyperparameter-free balance on multiple losses. In GSMorph, we reformulate the optimization procedure by projecting the gradient of similarity loss orthogonally to the plane associated with the smoothness constraint, rather than additionally introducing a hyperparameter to balance these two competing terms. Furthermore, our method is model-agnostic and can be merged into any deep registration network without introducing extra parameters or slowing down inference. In this study, We compared our method with state-of-the-art (SOTA) deformable registration approaches over two publicly available cardiac MRI datasets. GSMorph proves superior to five SOTA learning-based registration models and two conventional registration techniques, SyN and Demons, on both registration accuracy and smoothness.
\keywords{Medical image registration \and Gradient surgery \and Regularization.}
\end{abstract}
\section{Introduction}
Image registration is fundamental to many medical image analysis applications, e.g., motion tracking, atlas construction, and disease diagnosis~\cite{chen2021deep}. Conventional registration methods usually require computationally expensive iterative optimization, making it inefficient in clinical practice~\cite{avants2008symmetric,vercauteren2009diffeomorphic}. Deep learning has recently been widely exploited in the registration domain due to its superior representation extraction capability and fast inference speed~\cite{balakrishnan2018unsupervised,dalca2018unsupervised}. Deep-learning-based registration (DLR) formulates registration as a network learning process minimizing a composite objective function comprising one similarity loss to penalize the difference in the appearance of the image pair, and a regularization term to ensure the smoothness of deformation field. Typically, to balance the registration accuracy and smoothness of the deformation field, a hyperparameter is introduced in the objective function. However, performing hyperparameter tuning is labor-intensive, time-consuming, and \emph{ad-hoc}; searching for the optimal parameter setting requires extensive ablation studies and hence training tens of models and establishing a reasonable parameter search space. Therefore, alleviating, even circumventing, hyperparameter search to accelerate development and deployment of DLR models remains challenging.

Recent advances~\cite{chen2021registration,huang2021difficulty,jia2021learning} in DLR have primarily focused on network architecture design to boost registration performance. 
Few studies~\cite{hoopes2021hypermorph,mok2021conditional} investigated the potential in preventing hyperparameter searching by hypernetwork~\cite{ha2016hypernetworks} and conditional learning~\cite{huang2017arbitrary}. Hoopes~\textit{et al.}~\cite{hoopes2021hypermorph} leveraged a hyper-network that takes the hyperparameter as input to generate the weight of the DLR network. Although effective, it introduces a large number of additional parameters to the basic DLR network, making the framework computationally expensive. In parallel, Mok~\textit{et al.}~\cite{mok2021conditional} proposed to learn the effect of the hyperparameter and condition it on the feature statistics (usually illustrated as \textit{style} in computer vision~\cite{huang2017arbitrary}) to manipulate the smoothness of the deformation field in the inference phase. Both methods can avoid hyperparameter tuning while training the DLR model. However, they still require a reasonable sampling space and strategy of the hyperparameter, which can be empirically dependent.

Gradient surgery (GS) projects conflicting gradients of different losses during the optimization process of the model to mitigate gradient interference. This has proven useful in multi-task learning~\cite{yu2020gradient} and domain generalization~\cite{mansilla2021domain}. Motivated by these studies, we propose utilizing the GS to moderate the discordance between the similarity loss and regularization loss. The proposed method can further avert searching the weight for balancing losses in training the DLR.
\begin{itemize}
    \item We propose GSMorph, a gradient-surgery-based DLR model. 
    Our method can circumvent tuning the hyperparameter in composite loss function with a gradient-level reformulation to reach the trade-off between registration accuracy and smoothness of the deformation field. 
    \item Existing GS approaches have operated the parameters' gradients independently or integrally. 
    We propose a layer-wise GS to group by the parameters for optimization to ensure the flexibility and robustness of the optimization process. 
    \item Our method is model-agnostic and can be integrated into any DLR network without extra parameters or losing inference speed.
\end{itemize}

\section{Methodology}
\begin{figure*}[!htbp]
	\centering
	\includegraphics[width=0.9\linewidth]{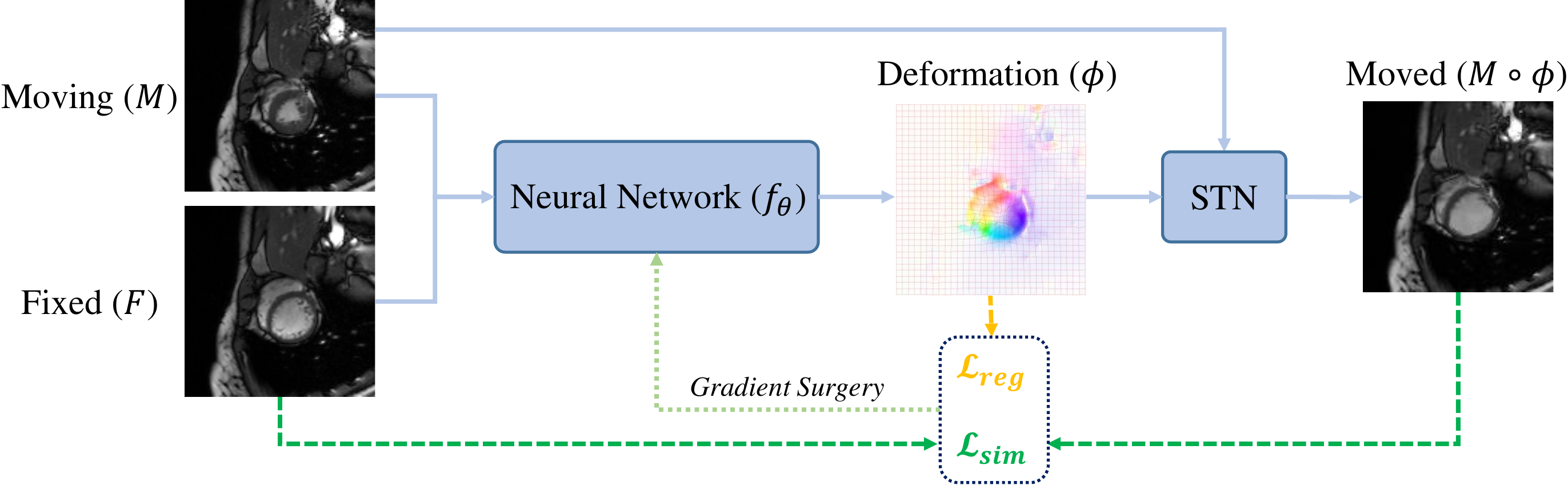}	
	\caption{Schematic illustration of our proposed GSMorph. GS modifies the gradients computed by similarity loss $\mathcal{L}_{sim}$ and regularization loss $\mathcal{L}_{reg}$, then updates the model's parameters {$\theta$}.}
	\label{fig:framework}
\end{figure*}

Deformable image registration estimates the non-linear correspondence field $\phi$ between the moving, $M$, and fixed, $F$, images (Fig.~\ref{fig:framework}). Such procedure is mathematically formulated as $\phi = f_\theta(F, M)$. For learning-based registration methods, $f_\theta$ (usually adopted by a neural network) takes the fixed and moving image pair as input and outputs the deformation field via the optimal parameters $\theta$. Typically, $\theta$ can be updated using standard mini-batch gradient descent as follows:
\begin{equation}
    \label{eq:GD}
    \theta := \theta - \alpha \nabla_\theta \left(\mathcal{L}_{sim}(\theta; F, M \circ \phi) + \lambda \mathcal{L}_{Reg}(\theta; \phi) \right)
\end{equation}
where $\alpha$ is the learning rate; $\mathcal{L}_{sim}$ is the similarity loss to penalize differences in the appearance of the moving and fixed images (e.g., mean square error, mutual information or local negative cross-correlation); $\mathcal{L}_{reg}$ is the regularization loss to encourage the smoothness of the deformation field (this can be computed by the gradient of the deformation field); $\lambda$ is the hyperparameter balancing the trade-off between $\mathcal{L}_{sim}$ and $\mathcal{L}_{reg}$ to achieve desired registration accuracy while preserving the smoothness of the deformation field in the meantime. 
However, hyperparameter tuning is time-consuming and highly experience-dependent, making it tough to reach the optimal solution. 

Insight into the optimization procedure in Eq.~\ref{eq:GD}, as registration accuracy and spatial smoothness are potentially controversial in model optimization, the two constraints might have different directions and strengths while going through the gradient descent. 
Based on this, we provide a geometric view to depict the gradient changes for $\theta$ based on the \textit{gradient surgery} technique. The conflicting relationship between two controversial constraints can be geometrically projected as orthogonal vectors. Depending on the orthogonal relationship, merely updating the gradients of the similarity loss would automatically associate with the updates of the regularization term. In this way, we avoid tuning the hyperparameter $\lambda$ to optimize $\theta$.
The Eq.~\ref{eq:GD} can then be rewritten into a non-hyperparameter pattern:
\begin{equation}
    \theta := \theta - \alpha \Phi ( \nabla_\theta \mathcal{L}_{sim}(\theta; F, M \circ \phi) )
\end{equation}
where $\Phi(\cdot)$ is the operation of proposed GS method.

\subsection{Layer-wise Gradient Surgery}
\begin{figure*}[!htbp]
	\centering
	\includegraphics[width=0.8\linewidth]{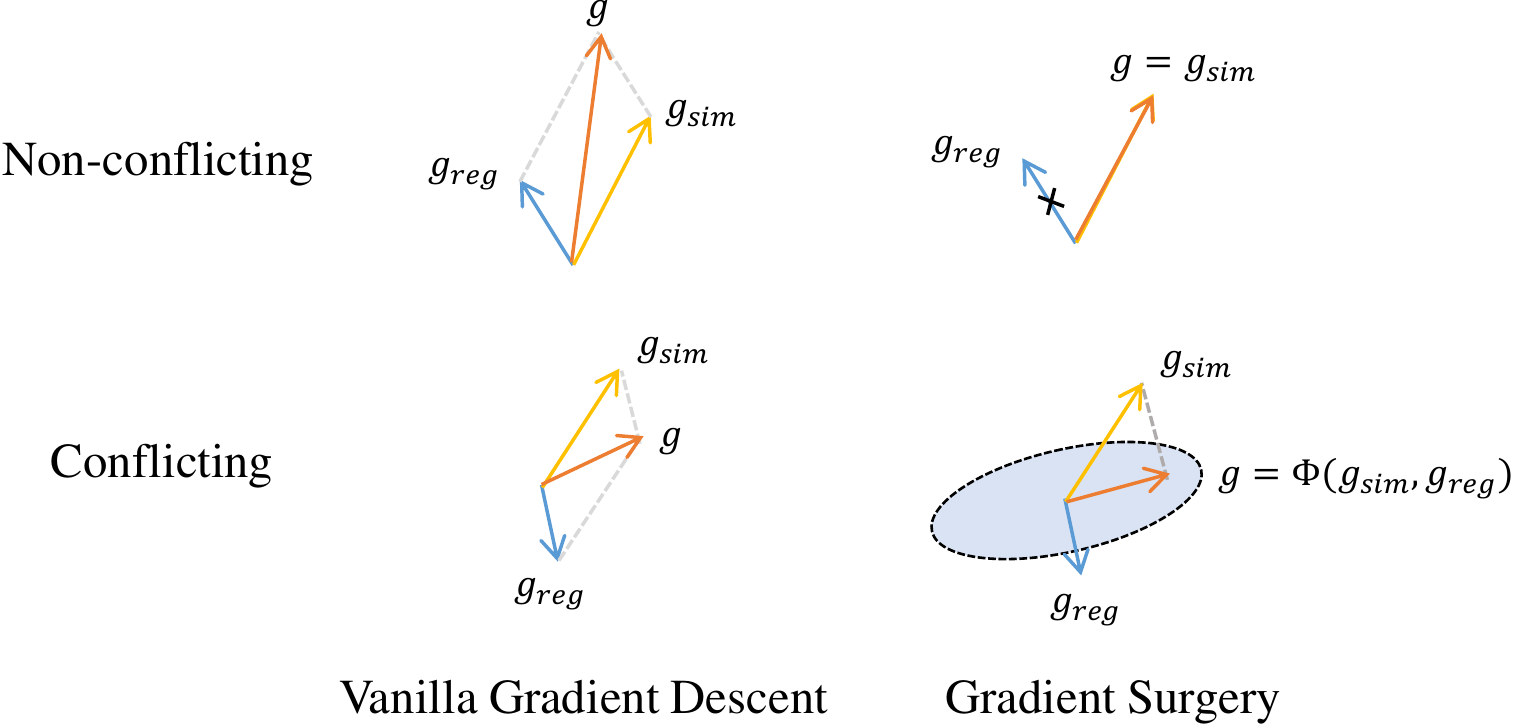}
	\caption{Visualization of vanilla gradient descent and gradient surgery for non-conflicting and conflicting gradients. Regarding vanilla gradient descent, the gradient, $g$, is computed based on the average of $g_{sim}$ and $g_{reg}$. Our GS-based approach projects the $g_{sim}$ onto the normal vector of $g_{reg}$ to prevent disagreements between the similarity loss and regularization loss. On the other hand, we only update the $g_{sim}$ in non-conflicting scenarios.
 }
\label{fig:gradientsurgery}
\end{figure*}

Figure~\ref{fig:gradientsurgery} illustrates the two scenarios of gradients while optimizing the DLR network via vanilla gradient descent or gradient surgery. We first define that the gradient of similarity loss, $g_{sim}$, and that of regularization loss, $g_{reg}$, are conflicting when the angle between $g_{sim}$ and $g_{reg}$ is the obtuse angle, viz. $\langle g_{sim}, g_{reg} \rangle < 0$. In this study, we propose updating the parameters of neural networks by the original $g_{sim}$ independently, when $g_{sim}$ and $g_{reg}$ are non-conflicting, representing $g_{sim}$ has no incompatible component of the gradient along the direction of $g_{reg}$. Consequently, optimization with sole $g_{sim}$ within a non-conflicting scenario can inherently facilitate the spatial smoothness of deformations.

Conversely, as shown in Fig.~\ref{fig:gradientsurgery}, conflicting gradients are the dominant reason associated with non-smooth deformations. Hence, deconflicting gradients in the optimization of the DLR network to ensure high registration accuracy, as well as smooth deformation, is the primary goal of our study. Following a simple and intuitive procedure, we project the $g_{sim}$ onto the normal plane of the $g_{reg}$, where the projected similarity gradient $g$ and $g_{reg}$ are non-conflicting along each gradient's direction. 

Existing studies~\cite{mansilla2021domain,yu2020gradient} performed the GS in terms of independent parameters or the entire network. 
Despite the effectiveness, these can be either unstable or inflexible. Considering that a neural network usually extracts features through the collaboration of each parameter group in the convolution layers, we introduce a layer-wise GS to ensure its stability and flexibility. The parameter updating rule is detailed in the Algorithm~\ref{alg:gradient surgery}. Specifically, in each gradient updating iteration, we first compute the gradients of two losses for the parameter group in each layer separately. Then, the conflicting relationship between the two gradients is calculated based on their inner production. Once the two gradients are non-conflicting, the gradients used to update its corresponding parameter group will be only the original gradients of similarity loss; on the contrary, the gradients will be the projected similarity gradients orthogonal to the gradients of regularization, which can be calculated as $g_{sim}^{i} - \frac{\langle g_{sim}^{i}, g_{reg}^{i} \rangle}{\| g_{reg}^{i} \|^2} g_{reg}^{i}$. After performing GS on all layer-wise parameter groups in the network, the final gradients will be used to update the model's parameters.
\begin{minipage}{.75\linewidth}
\centering
\begin{algorithm}[H]
\caption{Gradient surgery}
\label{alg:gradient surgery}
\begin{algorithmic}[1]
\Require Parameters $\theta_i$ in $i$th layer of the network; Number of layers in the network $N$. 
\State $g_{sim} \gets \nabla_\theta \mathcal{L}_{sim}$
\State $g_{reg} \gets \nabla_\theta \mathcal{L}_{reg}$
\For{$i = 1 \to N$}
\If{$\langle g_{sim}^{i}, g_{reg}^{i} \rangle > 0$}
    \State $g_i = g_{sim}^{i}$
\Else
    \State $g_i = g_{sim}^{i} - \frac{\langle g_{sim}^{i}, g_{reg}^{i} \rangle}{\| g_{reg}^{i} \|^2} g_{reg}^{i}$
\EndIf
\State $\Delta \theta_{i} = g_i$
\EndFor
\State Update $\theta$
\end{algorithmic}
\end{algorithm}
\end{minipage}
\subsection{Network Architecture}
Our network architecture (seen in Fig.~\ref{fig:framework}) is similar to VoxelMorph~\cite{dalca2018unsupervised} that comprises naive U-Net~\cite{ronneberger2015u} and spatial transform network (STN)~\cite{jaderberg2015spatial}. The U-Net takes the moving and fixed image pair as input and outputs the deformation field, which is used to warp the moving image via STN. The U-Net consists of an encoder and a decoder with skip connections, which forward the features from each layer in the encoder to the corresponding layer in the decoder by concatenation to enhance the feature aggregation and prevent gradient vanishing. The number of feature maps in the encoder part of the network is 16, 32, 64, 128, and 256, increasing the number of features as their size shrinks, and vice versa in the decoder part. Each convolutional block in the encoder and decoder has two sequential convolutions with a kernel size of 3, followed by a batch normalization and a leaky rectified linear unit. 

\section{Experiments and Results}
\subsection{Datasets and Implementations}
\subsubsection{Datasets}
In this study, we used two public cardiac cine-MRI datasets for investigation and comparison: ACDC~\cite{bernard2018deep} and M$\&$M~\cite{campello2021multi}. ACDC and M$\&$M contain 100 and 249 subjects, respectively. We followed a proportion of 75$\%$, 5$\%$, and 20$\%$ to split each dataset for training, validation, and testing. We selected the image from the cine-MRI cardiac sequence at the End Systole (ES) time point of the cardiac cycle as the moving image, and that at the End Diastole (ED) as the fixed one. All images were cropped into the size of 128$\times$128 centralized to the heart. We normalized the intensity of images into the range from 0 to 1 before inputting them into the model.

\subsubsection{Implementation details}
We implemented our model in PyTorch~\cite{paszke2019pytorch}, using a standard PC with an NVIDIA GTX 2080ti GPU. We trained the network through Adam optimizer~\cite{kingma2014adam} with a learning rate of 5e-3 and a batch size of 32 for 500 epochs. We also implemented and trained alternative methods for comparison with the same data and similar hyper-parameters for optimization. Our source code is available at \url{https://github.com/wulalago/GSMorph}.

\subsection{Alternative Methods and Evaluation Criteria}
To demonstrate the advantages of our proposed method in medical image registration, we compared it with two conventional deformable registration methods, i.e., \textbf{Demons}~\cite{vercauteren2009diffeomorphic} and \textbf{SyN}~\cite{avants2008symmetric}, and a widely-used DLR model, \textbf{VoxelMorph}~\cite{dalca2018unsupervised}. These methods usually need laborious effort in hyperparameter tuning. Additionally, we reported the results of VoxelMorph trained with different $\lambda$ (i.e., 0.1, 0.01, and 0.001, denoted as \textbf{VoxelMorph-l}, \textbf{VoxelMorph-m}, \textbf{VoxelMorph-s}). Meanwhile, we compared our approach to one alternative DLR model based on the hyperparameter learning, i.e., \textbf{HyperMorph}~\cite{hoopes2021hypermorph}. This method only require additional validations in searching the optimal hyperparameter without necessarily tuning it from scratch. Finally, we reformulated two variations of GS based on our concept for further comparison. Specifically, \textbf{GS-Agr}~\cite{mansilla2021domain} treats the gradient of each parameter independently. It updates the parameter with the gradient of similarity loss in the non-conflicting scenario, and a random gradient sampled from the Gaussian distribution when conflicting. While \textbf{GS-PCGrad}~\cite{yu2020gradient} uses the same GS strategy as ours, but with respect to the whole parameters of the entire network. The \textbf{Initial} represents the results without any deformation.

In this study, we used six criteria to evaluate the efficacy and efficiency of the investigated methods, including Dice score (Dice) and 95$\%$ Hausdorff distance (HD95) to validate the registration accuracy of the regions of interest, Mean square error (MSE) to evaluate the pixel-level appearance difference between the moved and fixed image-pairs, the percentage of pixels with negative Jacobian determinant (NJD) values to compare the smoothness and diffeomorphism of the deformation field, the number of parameters (Param) of the neural network and inference speed (Speed) to investigate the efficiency.

\subsection{Results}
As summarized in Table~\ref{tab:quantitative}, our method could obtain the best MSE in the ACDC dataset and Dice in the M$\&$M dataset while achieving comparable performance with the tuned VoxelMorph over other metrics.
The Dice and HD95 reported in Table~\ref{tab:quantitative} were averaged over three anatomical regions of interest in the heart, i.e., Left ventricle, Myocardium, and Right Ventricle (LV, Myo, and RV). Consequently, the proposed model achieved superior registration accuracy and spatial regularization with faster inference speed than the two conventional registration methods. We also observed that our approach gained higher registration performance than HyperMorph in both datasets. Regarding the GS-based methods, GS-Agr totally collapsed, as the conflicting gradients accounted for most have been replaced by random noise. On the other hand, GS-PCGrad only yielded an inadequate registration performance with an inclination of over-regularization. The comparison in the GS-based method shows the flexibility and robustness of our approach. 

\begin{table}[!htb]
\caption{Quantitative comparison of investigated methods on the testing datasets over ACDC and M$\&$M.}
\resizebox{\textwidth}{!}
{   
    \renewcommand{\arraystretch}{1.4}
    \centering
    \label{tab:quantitative}
    \begin{tabular}{c|cccc|cccc}
    \hline
    \hline
    \multicolumn{1}{c|}{\multirow{2}{*}{\diagbox{Methods}{Dataset}}}& \multicolumn{4}{c|}{\textbf{ACDC}} & \multicolumn{4}{c}{\textbf{M$\&$M}}\\\cline{2-9}
    \multicolumn{1}{c|}{} & Dice($\%$) &HD95(mm) &MSE($10^{-2}$) &NJD($\%$) & Dice($\%$) &HD95(mm) &MSE($10^{-2}$) &NJD($\%$) \\ 
    \hline
    Initial &61.81$\pm$8.68 &4.40$\pm$1.33  &1.58$\pm$0.52 & - & 61.03$\pm$10.16 & 4.79$\pm$1.82  & 1.90$\pm$1.08 & - \\
    \hline
    Demons &85.38$\pm$3.52 &1.67$\pm$0.75  &0.46$\pm$0.21 &1.31$\pm$0.59 & 75.66$\pm$10.30 & 17.79$\pm$6.23  & 0.71$\pm$0.61 & 1.84$\pm$1.19\\
    SyN &79.28$\pm$8.23 &2.24$\pm$1.28 &0.65$\pm$0.21 &0.30$\pm$0.27 & 81.97$\pm$9.36 &\textbf{2.45$\pm$2.04}  & 0.84$\pm$0.12 & 0.49$\pm$0.47 \\
    \hline 
    VoxelMorph-s &86.69$\pm$2.17 &1.30$\pm$0.24  &0.39$\pm$0.14 &2.01$\pm$0.96 &77.12$\pm$9.36 &3.43$\pm$2.18  &\textbf{0.42$\pm$0.29} & 3.45$\pm$2.33\\
    VoxelMorph-m &\textbf{87.47$\pm$2.21} &\textbf{1.29$\pm$0.30} &0.42$\pm$0.15 &0.67$\pm$0.48 &79.93$\pm$8.57 &2.91$\pm$1.98  &0.48$\pm$0.32 & 1.31$\pm$1.10\\
    VoxelMorph-l &82.12$\pm$4.30 &1.87$\pm$0.64  &0.59$\pm$0.18 &\textbf{0.10$\pm$0.14} & 77.18$\pm$8.69 & 2.81$\pm$1.60  & 0.74$\pm$0.43 &\textbf{0.16$\pm$0.22}\\
    \hline
    HyperMorph &83.44$\pm$3.55 &1.75$\pm$0.64  &0.47$\pm$0.20 &1.60$\pm$0.86 & 77.21$\pm$8.45 & 3.28$\pm$1.99  & 0.59$\pm$0.37 &2.50$\pm$1.22\\
    \hline
    GS-Agr &63.40$\pm$9.15 &4.20$\pm$1.35  &1.33$\pm$0.43 &0 &63.41$\pm$9.85 & 4.50$\pm$1.77  &1.55$\pm$0.86  & $<$0.001\\
    GS-PCGrad &84.59$\pm$3.53 &1.62$\pm$0.53  &0.51$\pm$0.16 &0.11$\pm$0.17 & 80.67$\pm$8.18 &2.48$\pm$1.67  & 0.59$\pm$0.36 & 0.41$\pm$0.44 \\
    GSMorph &87.45$\pm$2.27 &1.34$\pm$0.40   &\textbf{0.31$\pm$0.11} &0.87$\pm$0.52 &\textbf{82.26$\pm$6.59} & 2.66$\pm$1.93  &0.49$\pm$0.27 & 0.98$\pm$0.84\\
    \hline
    \hline
    \end{tabular}
}
\end{table}

\begin{figure*}[!htbp]
	\centering
	\includegraphics[width=0.9\linewidth]{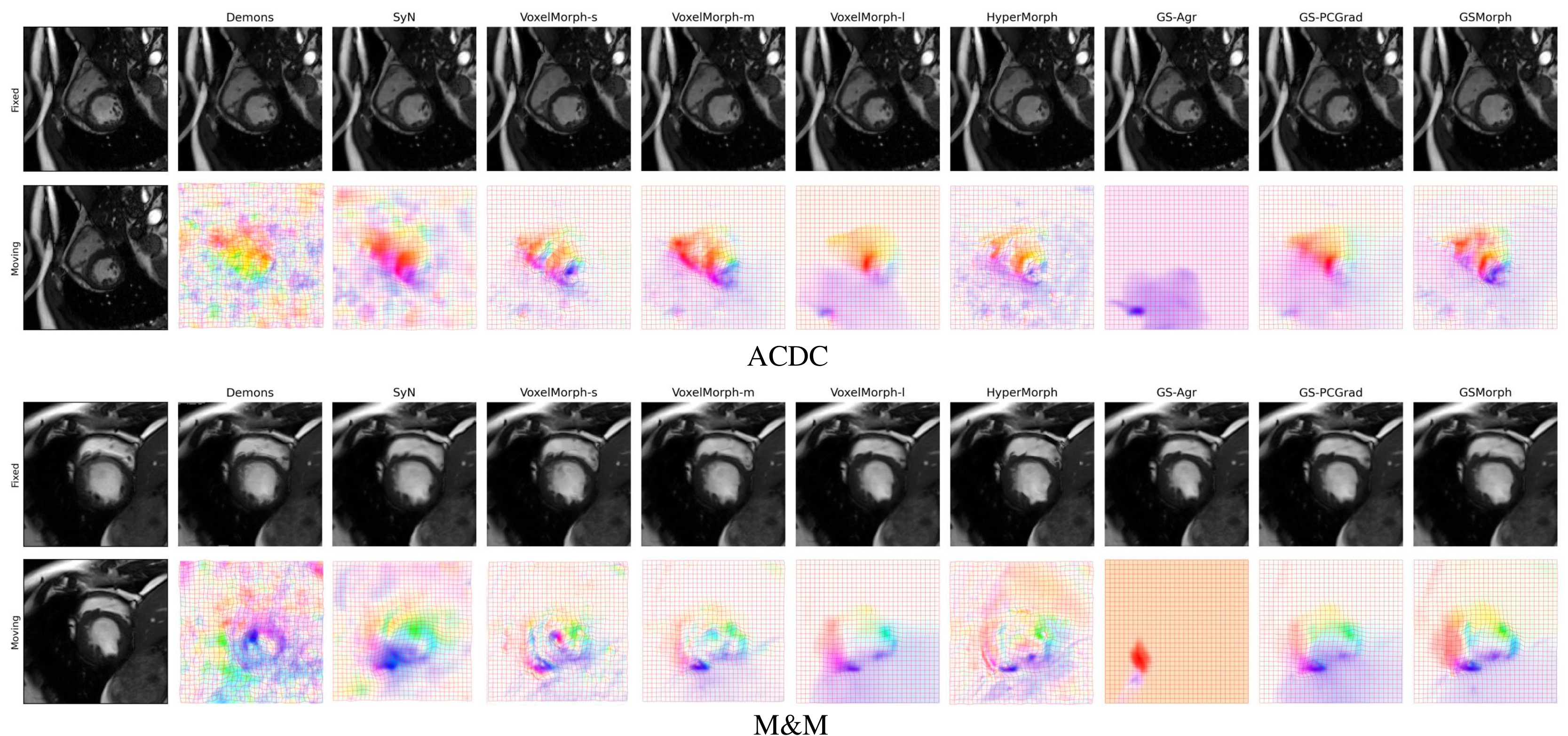}	
	\caption{Visual comparison of the registration results of the investigated methods for two representative test cases in ACDC and M$\&$M datasets. The top rows are the fixed images and moved images from different methods; the bottom rows are the moving images and deformation fields. (We encourage you to zoom in for better visualization)}
	\label{fig:visual}
\end{figure*}

Figure~\ref{fig:visual} illustrates the sample cases of the warped images and the corresponding deformation fields from the compared methods. It can be observed our methods could obtain the moved images most similar to the fixed ones. Voxelmorph could achieve comparable results to us but still require a time-consuming hyperparameter tuning.
Overall, the results of the comparisons in Table~\ref{tab:quantitative} and Fig.~\ref{fig:visual} indicate that our method performed the best among all the techniques that we implemented and examined, showing the effectiveness of our model in balancing the trade-off between registration accuracy and smoothness of deformations.

\begin{table}[!htb]
\caption{Number of parameters and inference speed of investigated methods on the testing datasets over ACDC and M$\&$M.}
\centering
\resizebox{0.8\textwidth}{!}
{   
    \renewcommand{\arraystretch}{1.4}
    \footnotesize
    \label{tab:param}
    \begin{tabular}{c|c|c|c|c|c}
    \hline
    \hline
        Methods & Demons & SyN & VoxelMorph & HyperMorph & GSMorph\\
        \hline
        Params & - & - & 1.96M & 126M & 1.96M \\
        Speed & 7.55$\pm$1.79 & 16.59$\pm$5.48 & 2.29$\pm$0.83 & 2.96$\pm$1.09 & 2.29$\pm$0.83 \\
    \hline
    \hline
    \end{tabular}
}
\end{table}

In Table~\ref{tab:param}, we have also reported the number of parameters and inference speed. We observed that DLR methods could obtain faster speed compared with conventional ones in general. As our proposed approach only modified the optimization procedure of the backbone network, it could maintain the original inference speed and the number of parameters. Conversely, HyperMorph introduced tremendous extra parameters and loss of inference speed as they adopted the secondary network to generate the conditions or weights of the main network architecture.

\section{Conclusion}
This work presents a gradient-surgery-based registration framework for medical images. 
To the best of our knowledge, this is the first study to employ gradient surgery to refine the optimization procedure in learning the deformation fields. In our GSMorph, the gradients from the similarity constraint were projected onto the plane orthogonal to those from the regularization term. In this way, merely updating the gradients in optimizing the registration accuracy would result in a joint updating of the gradients from the similarity and regularity constraints. 
Then, no additional regularization loss is required in the network optimization and no hyperparameter is further required to explicitly trade off between registration accuracy and spatial smoothness. Our model outperformed the conventional registration methods and the alternative DLR models. Finally, the proposed method is model-agnostic and can be integrated into any DLR network without introducing extra parameters or compromising the inference speed. 
We believe GSMorph will facilitate the development and deployment of DLR models and alleviate the influence of hyperparameters on performance.

\section{Acknowledgement}
This work was supported by the National Natural Science Foundation of China (62101365, 62171290, 62101343), Shenzhen-Hong Kong Joint Research Program (SGDX20201103095613036), Shenzhen Science and Technology Innovations Committee (20200812143441001), the startup foundation of Nanjing University of Information Science and Technology, the Ph.D. foundation for Innovation and Entrepreneurship in Jiangsu Province, the Royal Academy of Engineering (INSILEX CiET1819/19), Engineering and Physical Sciences Research Council UKRI Frontier Research Guarantee Programmes (INSILICO, EP/Y030494/1), and the Royal Society Exchange Programme CROSSLINK  IES$\backslash$NSFC$\backslash$201380.
%
%
%
\bibliographystyle{splncs04}
\bibliography{ref}
%

\end{document}